\documentclass[12pt]{article}
\usepackage{graphicx,epsfig}
\usepackage{amsmath,amssymb}
\def\to{\rightarrow}

\begin{document}
\begin{titlepage}

\begin{center}
\Large{Some Phenomenologies of Unparticle Physics} \vskip 1cm

\normalsize {Mingxing Luo\footnote{Email: luo@zimp.zju.edu.cn} and Guohuai Zhu\footnote{Email: zhugh@zju.edu.cn} \\
\vskip .5cm
 {Zhejiang Institute of Modern Physics, Department of Physics, \\
 Zhejiang University, Hangzhou, Zhejiang 310027, P.R. China}
  \\
 }
\begin{abstract}
Fermionic unparticles are introduced and their basic properties are discussed.
Some phenomenologies related are exploited,
such as their effects on charged Higgs boson decays and anomalous magnetic moments of leptons.
Also, it has been found that measurements of $B^0-\bar B^0$ mixing could yield interesting constraints
on couplings between unparticle operators and standard model fields.
\end{abstract}
\vskip 1cm
\end{center}

\end{titlepage}

Conformal invariance is a rarity in four dimensional quantum field theories for particle physics.
In general, conformal invariance will be broken by renormalization group effects
even if it exists classically.
Exceptional examples are the $N=4$ super-Yang-Mills theories,
which provide a rich laboratory for theoretical experimentation but have little relevance to real world phenomenologies.
Another type of examples are certain gauge theories,
which were first analyzed by Banks-Zaks ($\cal BZ$) many years ago \cite{bz}.
With suitable number of massless fermions,
these theories have non-trivial infrared fixed points,
which ensure a non-trivial conformal sector at the low energy limit.

Recently, it has been suggested \cite{g1,g2} that a conformal sector due to $\cal BZ$ fields
dubbed as unparticle physics might appear at the TeV scale.
If this is the case, things would change drastically and
phenomenologically relevant conformal invariance would then be awaiting us around the corner.
Things as such definitely have very distinct phenomenologies \cite{g1,g2,cky},
though there are many unsettled theoretical issues to be worked out.
For example, it is well known that $S$-matrices cannot be defined for conformal field theories,
as one cannot define asymptotic states in these theories.
Naively, on the other hand,
conformal invariance could be violated by couplings between the unparticles and the standard model particles.
It is argued in \cite{g1} that such couplings will not affect the infrared scale invariance of the unparticles,
as $\cal BZ$ fields decouple from ordinary matter at lower energy scale.
But it is still unclear how this can be consistently implemented in the framework of effective field theory.
Plus, general principles as guide-lines are still wanting to make realistic models.

Nevertheless, one may as well take such a novel framework as a working hypothesis.
One then pushes forward to see how far it can take us.
In this short note, fermionic unparticles are introduced and their basic properties are discussed.
Then, elementary phenomenologies related to unparticles are exploited.

To fix notations, we will start with the scalar unparticle operator $O_{\cal U}$.
Following \cite{g1}, due to scale invariance, one has
\begin{equation}
|\langle 0 |O_{\cal U} |P\rangle |^2 \rho(P^2) = A_{d_{\cal U}} \theta(P^0)\theta(P^2) (P^2)^{d_{\cal U} -2}
\label{s1}
\end{equation}
where
\begin{equation}
A_{d_{\cal U}} = {16\pi^{5/2}\over (2\pi)^{2d_{\cal U}}}
 {\Gamma(d_{\cal U}+1/2) \over \Gamma(d_{\cal U}-1) \Gamma(2d_{\cal U})}.
\end{equation}
Here $d_{\cal U}$ is the scale dimension of the operator $O_{\cal U}$
and eq. (\ref{s1}) can be interpreted sort of as a phase space of $d_{\cal U}$ massless particles.
This in turn yields the propagator for scalar unparticles:
\begin{equation}
\int d^4 x e^{iP\cdot x} \langle 0 |T O_{\cal U}(x) O_{\cal U}(0)  |0\rangle =
{i A_{d_{\cal U}} \over 2\sin (d_{\cal U}\pi)} {1\over (-P^2-i\epsilon)^{2-d_{\cal U}}}
\end{equation}

Similarly, for vector unparticles operator $O_{\cal U}^\mu$, one has \cite{g1,g2}
\begin{equation}
\langle 0 |O_{\cal U}^\mu |P\rangle  \langle P |O_{\cal U}^\nu |0\rangle  \rho(P^2)
= A_{d_{\cal U}} \theta(P^0)\theta(P^2) (P^2)^{d_{\cal U}-2 }
\left(-g^{\mu\nu} + {\xi P^\mu P^\nu\over P^2}   \right)
\end{equation}
and
\begin{equation}
\int d^4 x e^{iP\cdot x} \langle 0 |T O_{\cal U}^\mu(x) O_{\cal U}^\nu(0)  |0\rangle =
{i A_{d_{\cal U}} \over 2\sin (d_{\cal U}\pi)} {-g^{\mu\nu} + \xi P^\mu P^\nu/P^2 \over (-P^2-i\epsilon)^{2-d_{\cal U}}}
\end{equation}
respectively.
If one makes the extra assumption $\partial_\mu Q^\mu_{\cal U}=0$ (which is not necessary a priori), $\xi=1$.

So far, only the bosonic sector of unparticle physics is discussed.
We now make the extension to the fermionic domain.
We introduce unparticle operators $\Psi_{\cal U}$ which transform as spinors under Lorentz transformations.
Since the underlying theory is still assumed to be local quantum field theory,
the spin-statistics theorem still holds.
Thus, $\Psi_{\cal U}$ should obey anti-commutation rules, similar to ordinary fermions.

Parallel to bosonic cases, one defines,
\begin{equation}
\rho_{\alpha\beta}(P) = \langle 0 |\Psi^{\cal U}_\alpha |P\rangle \langle P |\bar\Psi^{\cal U}_\beta |0\rangle  \rho(P^2)
\end{equation}
$\rho(P^2)$ is a $4\times4$ matrix and can be expanded in terms of the 16 linearly independent products of the
$\gamma$ matrices: $1$, $\gamma^\mu$, $\sigma^{\mu\nu}$, $\gamma^5$, and $\gamma^\mu\gamma^5$.
The requirement of Lorentz covariance yields \cite{bd}
\begin{equation}
\rho_{\alpha\beta}(P) = \rho_1(P^2) \not P_{\alpha\beta} + \rho_2(P^2) \delta_{\alpha\beta}
+ \tilde\rho_1(P^2) (\not P\gamma^5)_{\alpha\beta} + \tilde\rho_2(P^2) i \gamma^5_{\alpha\beta}
\end{equation}
It is easy to prove on general ground that
(i) $\rho_1$, $\rho_2$, $\tilde\rho_1$, and $\tilde\rho_2$ are all real;
(ii) $\rho_1 \geq 0$; and (iii) $\rho_1 \geq |\tilde\rho_1 |$.

On the other hand, conformal invariance ensures that,
\begin{equation}
\rho_{\alpha\beta}(P) = B_{d_{\cal U}} \theta(P^0)\theta(P^2) (P^2)^{d_{\cal U} -5/2}
\left[ \left( (1-\alpha \gamma^5)\not P \right )_{\alpha\beta} + \zeta (P^2)^{1/2} (1+\beta i\gamma^5)_{\alpha\beta} \right]
\end{equation}
Here $\alpha$, $\beta$, $\zeta$ are real constants, and $|\alpha| \leq 1$ according to (iii).
This is how far the combination of Lorentz covariance and scale invariance can take us.
For simplicity and to reproduce the result of a massless fermion
when $\alpha=\pm1$ in the limit of $d_{\cal U} \rightarrow 3/2$,
we will assume $\zeta = 0$ from now on. That is,
\begin{equation}
\langle 0 |\Psi^{\cal U}_\alpha |P\rangle \langle P |\bar\Psi^{\cal U}_\beta |0\rangle  \rho(P^2)
= B_{d_{\cal U}} \theta(P^0)\theta(P^2)(P^2)^{d_{\cal U}-5/2} [ \not P (1 + \alpha \gamma^5)] _{\alpha\beta}
\end{equation}
Of course, one can get $\alpha=0$ by invoking the invariance of parity, which will not be assumed here.
Assuming TCP invariance, the corresponding propagator is
\begin{equation}
\int d^4 x e^{iP\cdot x} \langle 0 |T \Psi_\alpha^{\cal U}(x) \bar\Psi_\beta^{\cal U}(0)  |0\rangle =
{B_{d_{\cal U}} \over 2 i \cos (d_{\cal U}\pi)}
{[ \not P (1 + \alpha \gamma^5)] _{\alpha\beta} \over (-P^2-i\epsilon)^{5/2-d_{\cal U}}}~.
\end{equation}
To reproduce the massless fermion propagator at $d_{\cal U} \rightarrow 3/2$, one simple choice for $B_{d_{\cal U}}$
could be
\begin{equation}
B_{d_{\cal U}}=A_{d_{\cal U}-1/2}={32\pi^{7/2}\over (2\pi)^{2d_{\cal U}}}
 {\Gamma(d_{\cal U}) \over \Gamma(d_{\cal U}-3/2) \Gamma(2d_{\cal U}-1)}~ \nonumber.
\end{equation}

Now we are ready for some phenomenology.
Taking as an example, we introduce the following low-energy effective interaction term
\begin{equation}
\frac{C_{\cal U}\Lambda_{\cal U}^{k+3/2-d_{\cal U}}}{M_{\cal U}^k} \bar\Psi_{\cal U}(1-\gamma_5)e h_c+h.c.~,
\label{l1}
\end{equation}
where $\Psi_{\cal U}$ is a spinor unparticle operator and $h_c$ a charged Higgs boson.
Note that lepton numbers are violated if $\Psi_{\cal U}$ is not assigned a lepton number $+1$,
but electric charge is conserved and terms in (\ref{l1}) can be appropriately expanded to accommodate the $SU(2)_L$ symmetry.
It is convenient to rewrite the interaction term in terms of a dimensionless parameter
\begin{equation}
\frac{\lambda}{\Lambda_{\cal U}^{d_{\cal U}-3/2}}\bar\Psi_{\cal U}(1-\gamma_5)e h_c +h.c.~,
\mbox{~~~~~~with~~~~~~} \lambda=\frac{C_{\cal
U}\Lambda_{\cal U}^{k}}{M_{\cal U}^k}~.
\end{equation}
This will lead to the decay of a charged Higgs into an electron plus
unparticles of scale dimension $d_{\cal U}$, of the spectrum,
\begin{equation}
\frac{d\Gamma}{d E_e}= \frac{\lambda^2 B_{d_{\cal
U}}(1-\alpha)}{\pi^2 } \left (\frac{m_h}{\Lambda_{\cal U} }
\right )^{2d_{\cal U}-3}
\frac{E_e^2\theta(m_h-2E_e)}{m_h^2(m_h-2E_e)^{5/2-d_{\cal
U}}}~,
\end{equation}
where the electron mass has been neglected.
To avoid a non-integrable singularity as $E_e \to m_h/2$ in the above differential decay rate,
the scale dimension $d_{\cal U}$ should be larger than 3/2 for spinor unparticles.
The shape of the differential decay rate has the simple form
\begin{equation}
\frac{1}{\Gamma}\frac{d\Gamma}{d E_e}=(4d_{\cal
U}^2-1)(d_{\cal U}-\frac{3}{2})(1-2E_e/m_h)^{d_{\cal
U}-5/2} E_e^2/m_h^3~,
\end{equation}
which is, by replacing $d_{\cal U} \to d_{\cal U}+1/2$, actually the same
as that of a top quark decay into a up quark plus scalar unparticles \cite{g1}.
This interaction will also contribute to the lepton anomalous magnetic moments,
which can be readily calculated,
\begin{equation}
g_l-2=\frac{-\lambda^2 B_{d_{\cal
U}}(1-\alpha)}{24\pi^2\cos(d_{\cal U}\pi)} \left
(\frac{m_h}{\Lambda_{\cal U}} \right )^{2d_{\cal U}-3}
\frac{m_l^2}{m_h^2} \Gamma(\frac{7}{2}-d_{\cal U})
\Gamma(d_{\cal U}+\frac{1}{2})~.
\end{equation}
It is clear that, to get a finite contribution to $g-2$, $d_{\cal U}$ here should be smaller than $7/2$.
Therefore for spinor unparticles,
the scale dimension should fall into the intervals $3/2<d_{\cal U}<5/2$ or $5/2<d_{\cal U}<7/2$.
Note also that the spinor unparticles contribution to $g-2$ contains an extra factor
$(m_l/m_h)^{5-2du}$, compared with contributions from scalar unparticles (Eq. (\ref{gm2scalar}), see below).
To have some quantitative feeling, we take the following inputs for illustration
\begin{equation}
\lambda=1~, \hspace*{1cm}\alpha=0~,\hspace*{1cm}
m_h=100~\mbox{GeV}~,\hspace*{1cm} \Lambda_{\cal U}=1~\mbox{TeV}
\label{spinor-input}
\end{equation}
and find its contribution to the muon anomalous magnetic moments to be
$-1.8 \times 10^{-10} (7.5 \times 10^{-14})$ with $d_{\cal U}=2 (3)$,
to be compared with the experimental results with the Standard Model(SM) contributions subtracted \cite{PDG}
\begin{equation}
(g_\mu-2)_{exp}-(g_\mu-2)_{SM}=44(20)\times 10^{-10}~.
\label{discrepency}
\end{equation}

\begin{figure}[htb]
\centerline{\includegraphics[width=8cm]{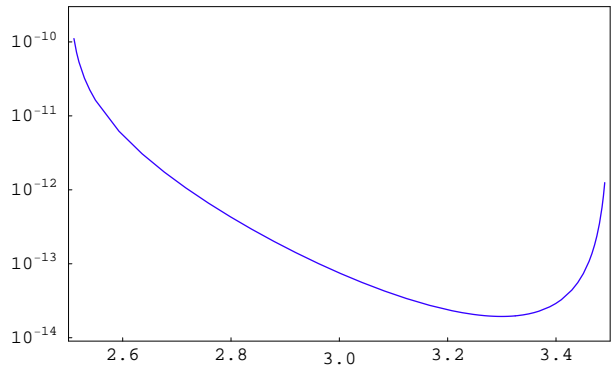}}
\caption{The fermionic unparticle contribution to $g_\mu-2$ as a function of scale dimension $d_{\cal U}$.}
\label{fig-gm2spinor}
\end{figure}

Actually, the spinor unparticle contribution to $g-2$ is always negative for $3/2<d_{\cal U}<5/2$,
which is opposite to the deviation of the SM predictions from experimental observations.
For $5/2<d_{\cal U}<7/2$, the spinor unparticle contribution does have the right sign, but its
magnitude is too small with inputs from Eq. (\ref{spinor-input}), as shown in Fig. \ref{fig-gm2spinor}.
Thus, such couplings with spinor unparticle operator seem not to provide an explanation for Eq. (\ref{discrepency}).

Usually, it is difficult to couple a single fermionic unparticle operator to SM particles,
as these couplings are strictly constrained by Lorentz invariance, gauge symmetries and other known discrete symmetries.
So, phenomenologies related with a single fermionic unparticle operator is relatively sterile compared with bosonic ones.
However, if one is willing to take two fermionic unparticle operators and to couple them with gauge bosons,
one gets much wider possibilities.
For example, one may include a term of the form,
\begin{equation}
\bar\Psi_{\cal U}^1 \gamma_\mu (1-\gamma^5) \Psi_{\cal U}^2 W^\mu
\end{equation}
which would contribute to $W$ decays. But such phenomena are probably hard to observe directly in experiments.

\begin{figure}[htb]
\centerline{\includegraphics[width=8cm]{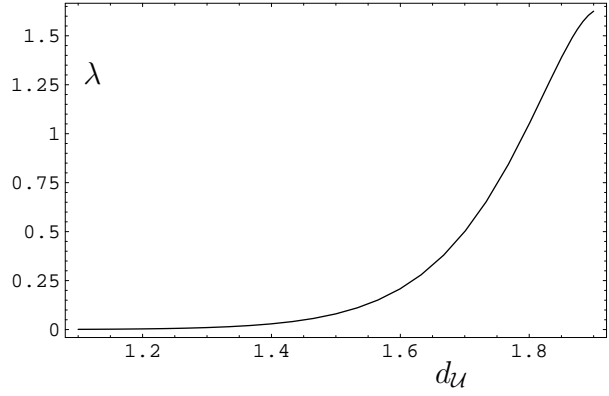}}
\caption{Assuming that the central value of Eq. (\ref{discrepency}) could be accounted for by scalar unparticles,
the coupling parameter $\lambda$ would be determined as a function of scale dimension $d_{\cal U}$. }
\label{fig-gm2scalar}
\end{figure}

Coming back to the bosonic sector, let us calculate the contribution of the following coupling of
scalar unparticle operator
\[
\frac{\lambda}{\Lambda_{\cal U}^{d_{\cal U}-1}} \bar{l}
O_{\cal U} l
\]
to the lepton anomalous magnetic moments:
\begin{equation}\label{gm2scalar}
g_l-2=\frac{-\lambda^2 A_{d_{\cal U}}}{8\pi^2\sin(d_{\cal
U}\pi)} \left (\frac{m_l}{\Lambda_{\cal U}} \right
)^{2d_{\cal U}-2} \left ( \frac{\Gamma(2-d_{\cal U})
\Gamma(2d_{\cal U}-1)}{\Gamma(1+d_{\cal U})}+
\frac{\Gamma(3-d_{\cal U}) \Gamma(2d_{\cal
U}-1)}{\Gamma(2+d_{\cal U})} \right )~.
\end{equation}
To get a finite result, the scale dimension $d_{\cal U}$ here should be smaller than $2$.
It can explicitly be checked that, at the limit $d_{\cal U} \to 1$,
Eq. (\ref{gm2scalar}) reproduces the higgs contribution to $g-2$ in the SM.
It can be seen from Fig. \ref{fig-gm2scalar} that, by taking $\Lambda_{\cal U}=1$ TeV, the deviation of the SM predictions from experimental observations
on $g_\mu-2$ does lead to some restrictions on the scalar unparticle operators, especially at lower $d_{\cal U}$ region.

Scalar and vector unparticle operators may in principle contribute to
flavor changing neutral current processes.
For example, the following effective interaction terms with scalar unparticle operator
may contribute to $B^0-\bar{B}^0$ mixing
\begin{equation}
\frac{\lambda}{\Lambda_{\cal U}^{d_{\cal U}}} \bar{d}
\gamma_\mu (1-\gamma_5) b \partial^\mu O_{\cal U}+h.c. ~.
\label{bbs}
\end{equation}
For simplicity, $\lambda$ here is assumed to be real.
Recall that in the basis of flavor eigenstates,  the time-evolution of the $B^0-\bar{B}^0$ system is determined
by the matrix
\begin{equation}
\hat{H}=\left ( \begin{array}{cc}
           M-i\frac{\Gamma}{2} & M_{12}-i\frac{\Gamma_{12}}{2} \\
           M_{12}^*-i\frac{\Gamma_{12}^*}{2}   &   M-i\frac{\Gamma}{2}
\end{array} \right )
\simeq \left ( \begin{array}{cc}
           M-i\frac{\Gamma}{2} & M_{12} \\
           M_{12}^*   &   M-i\frac{\Gamma}{2}
\end{array} \right )
\end{equation}
This approximation is justified since $\Gamma_{12} \ll M_{12}$.
When unparticle contributions are included, we have
\begin{equation}
\hat{H}=\left ( \begin{array}{cc}
           M-i\frac{\Gamma}{2} & M^{SM}_{12}+M^{\cal U} \\
           (M^{SM}_{12})^*+M^{\cal U}   &   M-i\frac{\Gamma}{2}
\end{array} \right )~.
\end{equation}
Notice that $M^{\cal U}$ does not contain weak phases for a real $\lambda$.
Since the SM can already account for the experimental observation
$\Delta M_d^{exp} =3.34 \times 10^{-13}~\mbox{GeV}$ \cite{PDG} within theoretical uncertainties,
$M^{\cal U}$ should be much smaller than $M^{SM}_{12}$.
We have thus
\begin{align}
\Delta M_d = \Delta M_d^{SM}+\Delta M_d^{\cal U} & = 2 Re \sqrt{(M^{SM}_{12}+M^{\cal U})((M^{SM}_{12})^*+M^{\cal U})} \\
          & \simeq 2 \vert M^{SM}_{12} \vert + 2 \cos (2\beta) Re (M^{\cal U})~,
\end{align}
where $\beta$ is one of the angles in the CKM triangle.
It is then straightforward to obtain the unparticle contributions to the mass difference of the neutral B mesons
\begin{equation}
\Delta M_d^{\cal U} = \frac{5\lambda^2 A_{d_{\cal U}}\cos(2\beta)
}{6} \left ( \frac{m_B}{\Lambda_{\cal
U}} \right )^{2d_{\cal U}} \frac{B_{B_d}f_{B_d}^2}{m_B}~,
\end{equation}
which should be significantly smaller than $\Delta M_d^{exp}$.

Similarly, unparticles could contribute to the width difference of neutral B mesons,
\begin{equation}
\Delta \Gamma_d^{\cal U} = 4 \cos (2\beta) Im (M^{\cal U})=
\frac{5\lambda^2 A_{d_{\cal U}}\cos(2\beta)\cos(d_{\cal U}\pi)}{3\sin(d_{\cal U}\pi)}
\left ( \frac{m_B}{\Lambda_{\cal U}} \right )^{2d_{\cal U}} \frac{B_{B_d}f_{B_d}^2}{m_B}~.
\end{equation}
Note that experimentally the width difference of the neutral B mesons has not been observed yet \cite{HFAG},
\begin{equation}
\frac{\Delta \Gamma_d}{\Gamma_d}=0.009 \pm 0.037~,
\end{equation}
here $\Gamma_d$ is the averaged decay width of neutral B mesons. Since the SM prediction on $\Delta \Gamma_d$
is very small, it seems reasonable to take the following upper limit for unparticle contributions,
\begin{equation}
\frac{\Delta \Gamma_d^{\cal U}}{\Gamma_d}<0.05~~ \Longrightarrow~~
\Delta \Gamma_d^{\cal U}<2.16 \times 10^{-14}~\mbox{GeV}~,
\end{equation}
which also provides constraint on unparticle coupling parameter.

\begin{figure}[htb]
\centerline{\includegraphics[width=8cm]{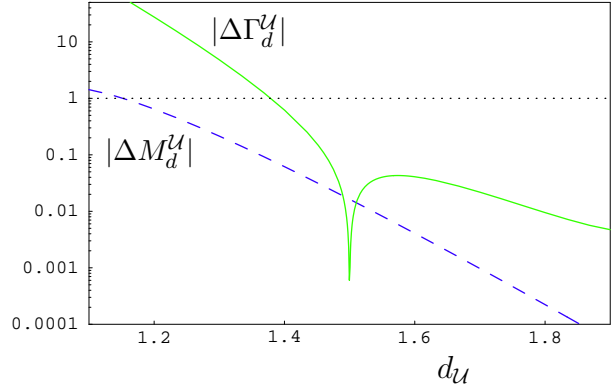}}
\caption{The scalar unparticle contribution to the mass and width differences of $B^0-\bar{B}^0$ system, normalized by
$\Delta M_d^{exp}$ and $2.16 \times 10^{-14}$GeV, respectively, are plotted as a function of scale dimension $d_{\cal U}$. }
\label{BBbar-mixing}
\end{figure}

For illustration, the mass and width differences $\Delta M_d^{\cal U}$ and $\Delta \Gamma_d^{\cal U}$ are plotted
as a function of $d_{\cal U}$ in Fig. \ref{BBbar-mixing}, by taking the following inputs
\begin{equation}
\sqrt{B_{B_d}}f_{B_d}=0.2~\mbox{GeV}~,\hspace*{1cm}
\Lambda_{\cal U}=1~ \mbox{TeV}~,\hspace*{1cm} \lambda=0.005~, \hspace*{1cm} \beta=21.2^\circ~
\label{mixing-input}
\end{equation}
Here the value of angle $\beta$ is quoted from \cite{HFAG}. It is clear that $B^0-\bar{B}^0$ mixing gives a
strong constraint on the unparticle coupling parameter $\lambda$, especially when the scale dimension
$d_{\cal U}$ is smaller than $1.4$.

Replacing down quark by strange quark, this effective operator will also contribute to the $B_s-\bar{B}_s$ mixing.
Since the mass difference $\Delta M_s=1.17 \times 10^{-11}~\mbox{GeV} $ \cite{CDF}
is about 30 times larger than $\Delta M_d$
but the effects of unparticles on them are roughly the same, it will give a milder
constraint on the coupling parameter $\lambda$.

Vector unparticles may also contribute to $B^0-\bar{B}^0$ mixing
through similar effective operators
\begin{equation}
\frac{\lambda}{\Lambda_{\cal U}^{d_{\cal U}-1}} \bar{d}
\gamma_\mu (1-\gamma_5) b O^\mu_{\cal U}+h.c. ~.
\end{equation}
It is easy to find that, the effect from the above operators on $\Delta M_d$ and $\Delta \Gamma_d$ are almost
the same as the case of scalar unparticles, except for an extra factor $(8/5-\xi)(\Lambda_{\cal U}/m_B)^2$.

Before closing, we would like to speculate the following, maybe wild, possibility.
Even though unparticle physics and supersymmetry are logically independent,
it may prove to be fruitful to combine them together in one framework.
Given the examples of $N=4$ super-Yang-Mills theories and
possible infrared fixed points in a variety classes of $N=1$ super gauge theories,
one may suspect that supersymmetry is one essential if not necessary ingredient to preserve conformal invariance.
This gives some rational for such a combination.
Technically, it is rather straightforward to do so.
For example, one can introduce chiral super-unparticle fields
\begin{equation}
\Phi_{\cal U} = \phi_{\cal U} + \sqrt{2} \theta \Psi_{\cal U} + \theta^2 F_{\cal U}
\end{equation}
and similar vector super-unparticle fields.
Upgrading every field in the SM into a superfield, one easily builds up supersymmetric couplings by the usual recipe.
Phenomenologies of such theories could be interesting.

{\bf Acknowledgements:} This work is supported in part by the National Science Foundation of China
 under grant No. 10425525 and No. 10645001.

\end{document}